\documentclass[prl,twocolumn,showpacs,amsfonts,amssymb]{revtex4}

\usepackage{graphicx}
\usepackage{amsmath}
\usepackage{bm}
\usepackage{times}
\usepackage{hyperref}

\newlength{\figwidth}
\renewcommand{\abstractname}{}\abstractname

\begin{document}
\title{Macroscopic magnetic guide for cold atoms} 
\author{Alexey Tonyushkin}
\email{alexey@physics.harvard.edu}
\author{Mara Prentiss}
\affiliation{Department of Physics, Harvard University, Cambridge, Massachusetts 02138, USA}
%\date{}
\begin{abstract}
We demonstrate a macroscopic magnetic guide for cold atoms with suppressed longitudinal field curvature which is highly desired for atom interferometry. The guide is based on macroscopic copper tape coils in a copropagating currents geometry, where the atoms are located between the coils few cm away from each surface.
The symmetric geometry  provides a much lower magnetic field curvature per fixed length that promises longer coherence time for atom interferometers. A double-tape design of each coil allows a smooth translation of guided atoms without addition of an external bias field. The guide is also immune from the current and thermal noise by virtue of the turns averaging and a large working distance, respectively. We present the experimental results of guide application to atom interferometry.
%Compared to a ferromagnetic guide, the new guides offer precise control over the magnetic field.
\end{abstract}
% 03.75.Be, 32.80.Pj,03.75.-b
\maketitle 
%%%%%%%%%%%%%%%%%%%%%%%%%%%%%%%%%%%%%%%%%%%%%%%%%%%%%%%%%%%%%%%%%%%%%%%%%%%%%%%%%%%%%%%%%

Presently, the most common type of magnetic guides for neutral atoms are micro-fabricated structures on dielectric substrates~\cite{achip}. Due to requirements on the potential smoothness, such a guide needs to be fabricated with submicron precision~\cite{noise}, which is still considered a challenge. Recently a partial solution to roughness suppression was demonstrated via rapid current modulation~\cite{modulation}. Roughness can also be reduced by addition of many wires to compensate for finite size~\cite{manywires}. The other disadvantage of the micro-chip is its close surface operation, which causes spin flips and thus atom loss~\cite{decoherence}. The short distance imparts the requirement for a chip to be placed inside a vacuum system and often integration of optical elements onto the chip itself thus limiting tuning capabilities of such systems~\cite{BECchip}. 
%The other drawback of such system is additional heating and interaction from the surface and therefore short coherence time.
One alternative to the micro-chip guide is an Ioffe-type four-rods guide frequently used as a conveyor to transport atoms along its guiding direction~\cite{4rods}. The major problem with such guides, however, is the requirement of a high current of a few hundred amps~\cite{Cren}, which requires water cooling, and the inability to smoothly translate atoms in the transverse direction. Another alternative is the use of an electromagnet-based guide with ferromagnetic materials~\cite{foils}.  
%This type of guide was used to demonstrate an enclosed area atom interferometer suitable for rotational sensing~\cite{Wu}.
Because of field enhancement in ferro-foils the operational distance can be large; however, the hysteretic nature of ferromagnetic materials does not allow precise control over the magnetic field.

The goal of this effort is to create a magnetic guide with the reduced potential roughness and small field curvature capable of smooth translation of the atoms in the transverse direction. A flat potential along the guiding direction is required to achieve a long interrogation time for atom interferometer-based sensors~\cite{sensors}. To date, none of the demonstrated finite-size guides for atoms provide both the flat longitudinal potential and high gradient in transverse direction. In this work we present the design of such macroscopic guiding structure that utilizes symmetric geometry. An alternative geometry with low intrinsic field curvature although with low field gradient and no radial access is provided by a solenoidal geometry~\cite{Opat}. Our guide is based on a parallel copropagating currents in a geometry similar to microscopic two-wire guides used as an atom conveyor~\cite{2wire}. 
In addition such a guide allows precise cancellation of magnetic field and provides initial trapping potential for a surface magneto-optical trap (MOT). It is easily fabricated and does not require water cooling to generate large field gradients. 

\begin{figure}[htb!]
\begin{center}
\includegraphics[width=2.6 in]{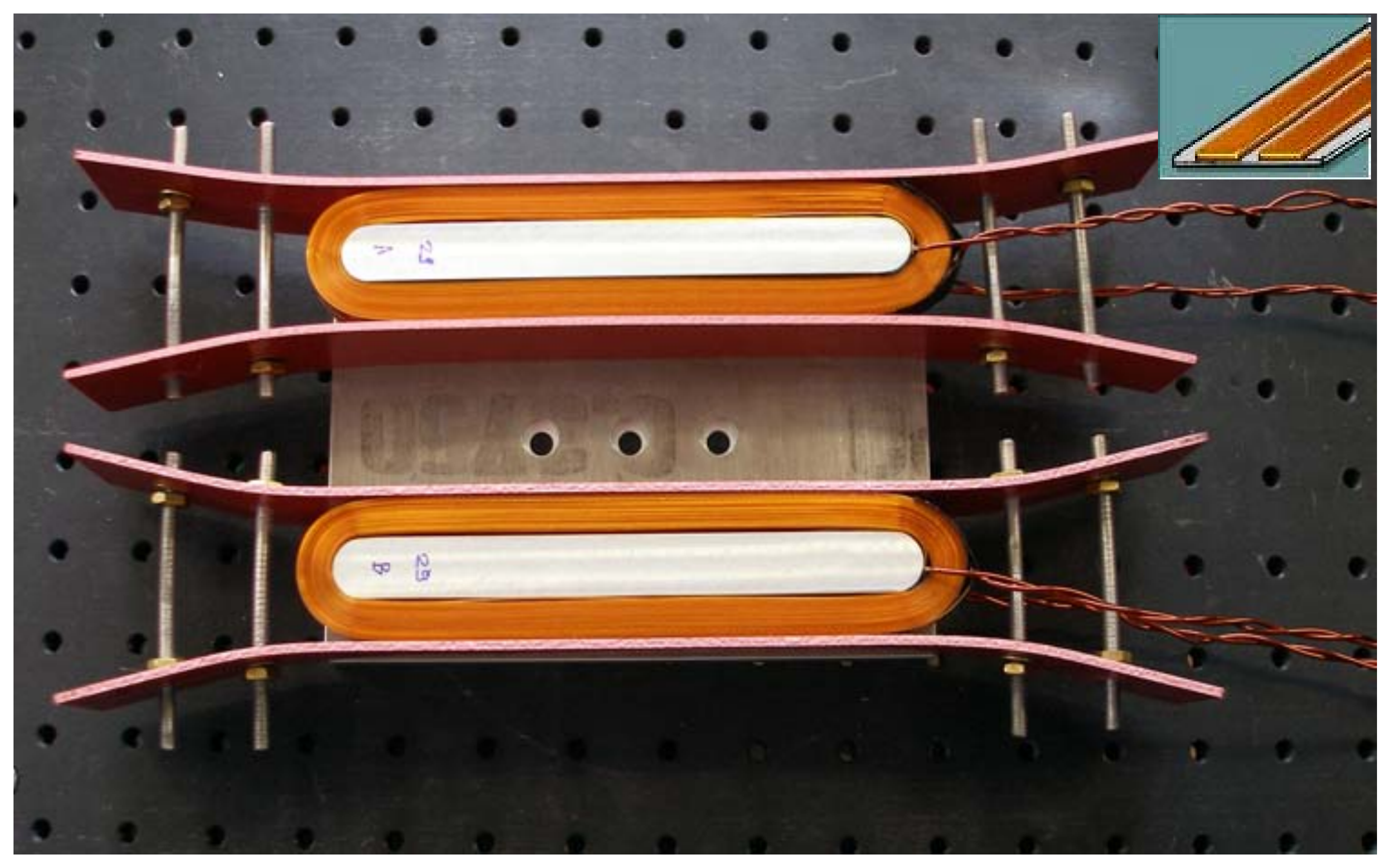}
\end{center}
\caption{(Color online) Photography of macroscopic symmetric magnetic guide structure, the inset picture shows the double tape conductor.}
\label{photo}
\end{figure}

The guiding structure developed here is based on coils with macroscopic copper tape (Bridgeport Magnetics) wrapped around aluminum structures as shown in Fig.~\ref{photo}.  
%%%%%%%%%%%%%%%%%%%%%%%%%%%%%%%%%%%%% 
%%%%%%%%%%%%%%%%%%%%%%%%%%%%%%%%%%%%%%%%%%%%%%%%%%%%%%%%%%%% 
Each coil has a 150 mm $\times$ 15 mm aluminum core and consists of 29 turns of double conductor tape (6.35 mm wide and 0.25 mm thick) with a 0.025-mm-thick layer of Kapton insulator on one side. The spacing between two coils $h$ can be varied to accommodate vacuum cell of different sizes. We enclosed each coil in glass-epoxy dielectric retainers to insure that the coils are straight and parallel to each other.
In the experiment, we place the guiding structure outside of an ultrahigh vacuum glass cell with square 
($4 \times 4$~cm) cross-section. The guide is mounted on two translational and two rotational stages that allow initial alignment of the guide axis with respect to the optical beams. In the stationary guide regime we use only one set of conductors from each coil, which are connected in series. Open air provides enough heat dissipation for the applied current of up to 50~A in the pulsed current regime. In the preliminary design we used similar coils in a counterpropagating current geometry to generate a minimum of the magnetic field above the coils plane.
That macro guide, however, suffered significantly from the potential variation associated with the respective current noise in two coils, which resulted in the short coherence time of the atom interferometer. 

\begin{figure}[htb!]
\begin{center}
\includegraphics[width=3.2 in]{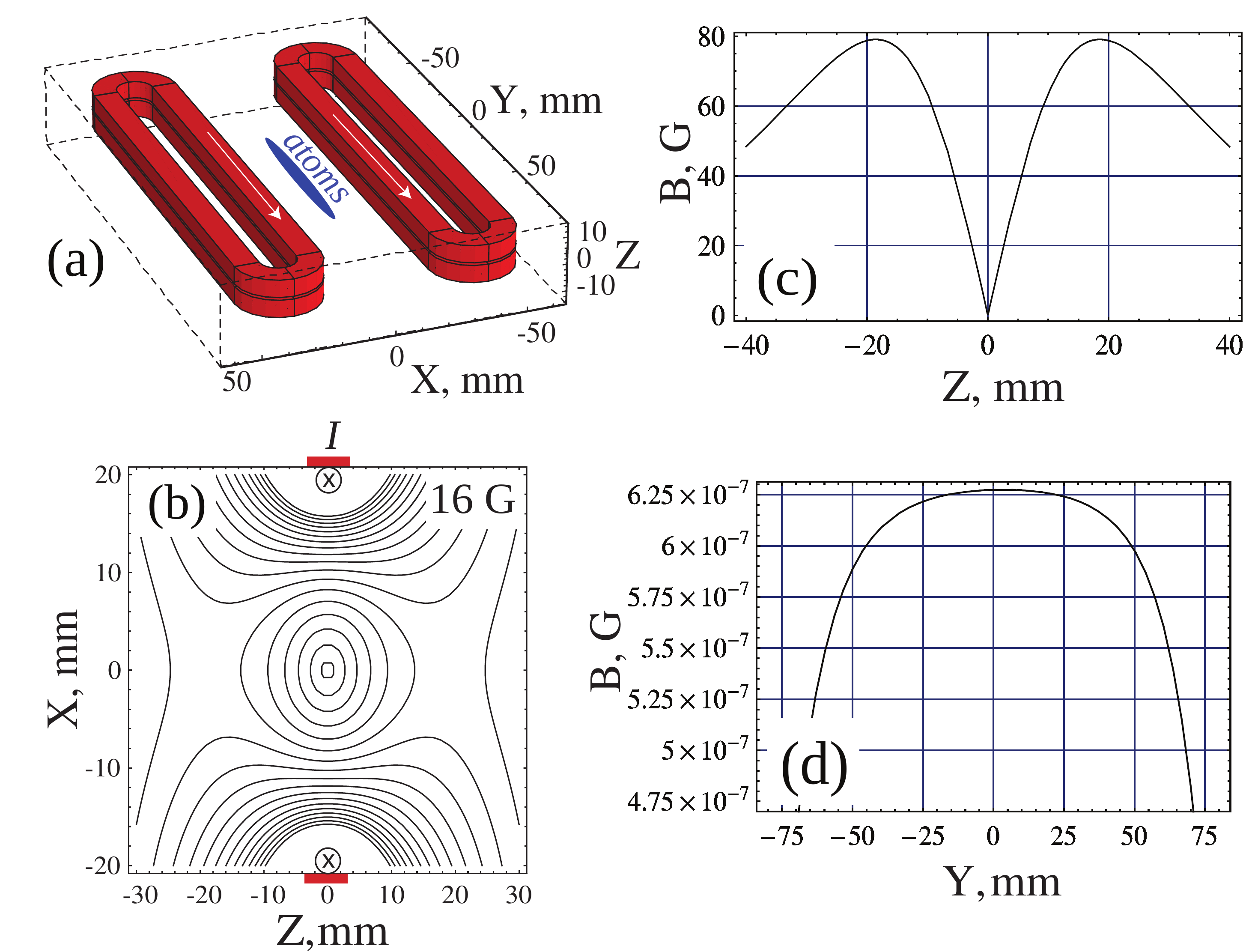}
\end{center}
\caption{(Color online) Symmetric guide configuration (a), and magnetic field of the guide for current of 50~A: (b) contour plot of the magnetic field vs position (x,z) (the separation between the lines is 16G, the tape conductors from two coils and direction of the current are shown for the reference); (c) magnetic field vs x; (d) magnetic field vs y, which gives the curvature of the field near the axis.}
\label{field}
\end{figure}

The magnetic field properties of the symmetric guide are shown in Fig.~\ref{field}. 
%%%%%%%%%%%%%%%%%%%%%%%%%%%%%%%%%%%%% Experiment scheme
%%%%%%%%%%%%%%%%%%%%%%%%%%%%%%%%%%%%%%%%%%%%%%%%%%%%%%%%%%%%
In our setup the quadrupole guiding potential is created along the symmetry axis between coils 2 cm away from the outer turn of each coil. A peak current $I=50$~A in the guiding regime provides a field gradient of 80~G/cm and a guide depth of $\approx 2.7$~mK for the $|F=1, m_{F}=-1>$ ground state of $^{87}$Rb. 
The magnetic field and radial gradient of the field close to the symmetry axis scale with coils separation as $\propto 1/h^2$.
% where $d$ is the separation between the (outer surfaces) coils.
%Since we use a relatively large glass cell and respectively large separation, we do not fully benefit from maximum field gradients of our guide. 
Microscopic separation of the coils can provide field gradients of the order of a few kG/cm,  
which are highly desirable for ultracold atom experiments in low dimensions~\cite{lowdim}. Another advantage of our guide is favorable field line orientation with respect to polarization of the optical molasses beams in the mirror MOT regime that results in the increased capture volume by a factor of five compared to the electromagnets used in counterpropagating current geometry. Figure~\ref{field}(d) shows the field variation near the symmetry axis along the guiding direction, the estimated curvature of the field $\partial ^2 B/\partial y^2 \approx 8.7 \times 10^{-8}$~G/cm$^{2}$, which is many orders of magnitude less than curvatures of the similar size guides in a counterpropagating current geometries.  

The typical experimental procedure is similar to the one described for a ferro-foils guide in Ref.~\cite{Wu}.
We first use the guiding coils in a low current mode ($I_1 = 6-8$~A, $\nabla B \sim 12-16$~G/cm) to capture and cool atoms that are optically transported from a source chamber along z-axis into a mirror MOT that is generated by four 45$^{\circ}$-oriented laser beams bounced from a mirror in the inner surface of the glass chamber. Atoms collected by the mirror MOT and Doppler-cooled for 500~ms, after which the optical molasses beams are gradually turned off and the atoms are optically pumped into a low-field-seeking hyperfine magnetic sublevel, at the same time the current in the coils is ramped up by the current regulated power supply to the peak value of the guide. We experimentally found that there is no need for the longitudinal bias field since there is always a residual offset field present. We estimate the transverse trapping frequency $\omega_{\perp} = 2\pi \times 40$~Hz with the atoms temperature $T \approx 30 \, \mu$K. The loading efficiency is maximized by monitoring the absorption of a weak near-resonant probe beam propagating along the guiding direction with the typical guide absorption of up to $90\%$ that corresponds to the optical depth of more than 20 along the guide. 

\begin{figure}[hb!]
\begin{center}
\includegraphics[width=\figwidth]{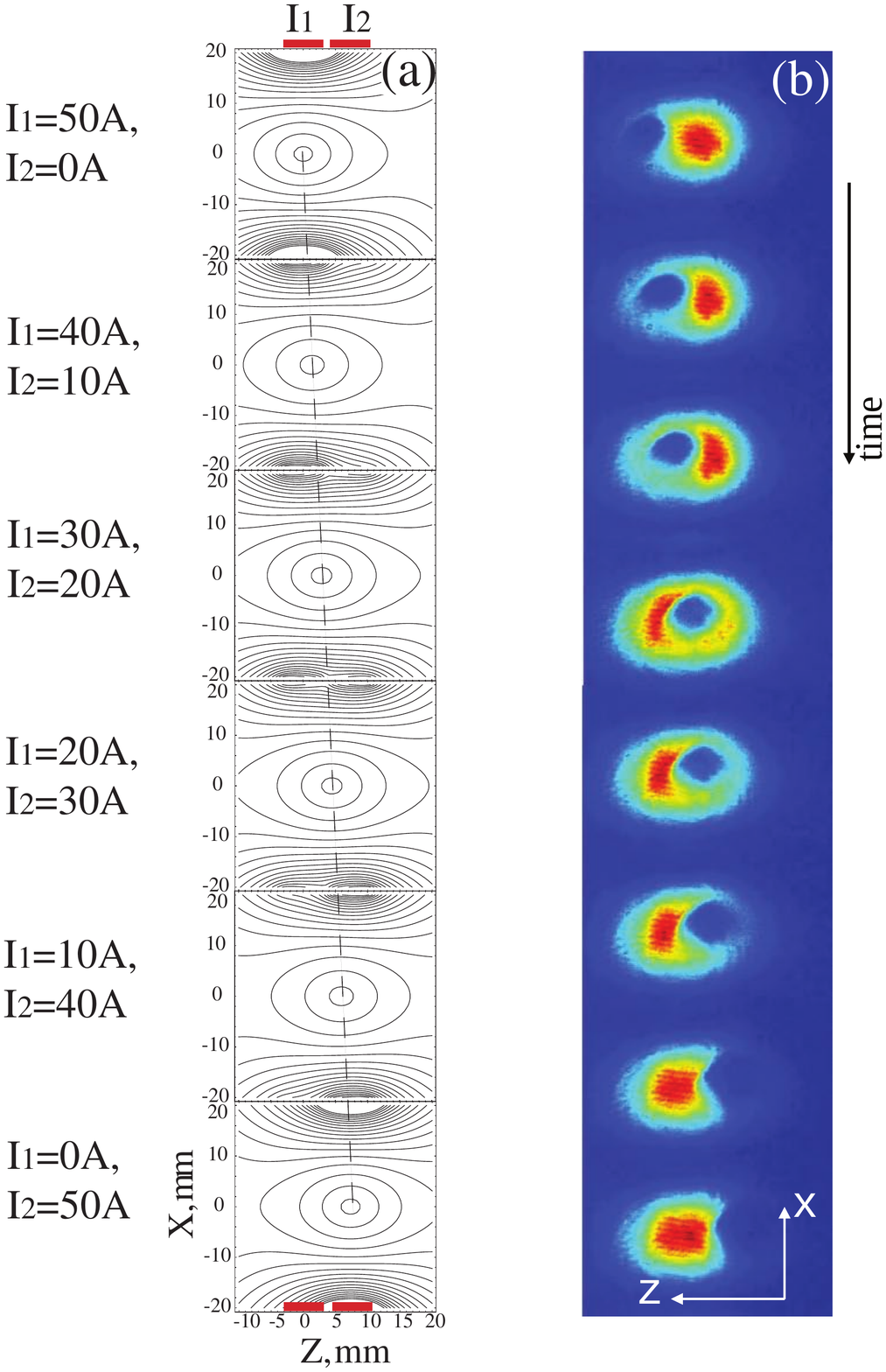}
\end{center}
\caption{(Color online) Guide translation: (a) simulations of the field contours show field minimum translation over 7 mm for $I_1 = 50-0\, A, I_2 = 0-50 \, A$ (the tape conductors are shown for reference); (b) experimental data of absorption images of guided atoms across the resonant probe beam; the guided atoms are smoothly translated over $\approx 2$~mm distance (from top to bottom) limited by the probe diameter.}
\label{moving}
\end{figure}

The main goal of our work is to develop a guide that allows us to enclose a large area in the atom-interferometer-based rotational sensor by translating atoms over the large distances~\cite{Wu}. In contrast to the early demonstration of the translated guide, our guide allows fast translation over cm distances. Figure~\ref{moving} shows translation of the guided atoms over such macroscopic distance along z-axis. 
%%%%%%%%%%%%%%%%%%%%%%%%%%%%%%%%%%%%% 
%%%%%%%%%%%%%%%%%%%%%%%%%%%%%%%%%%%%%%%%%%%%%%%%%%%%%%%%%%%% 
Unlike micro-chips used for atom transport above the surface, symmetry insures that the atoms do not experience bumps in the x-axis direction over the course of travel. 
To get the absorption images of guided atoms we direct a weak probe beam on-resonance to the 
$D_2 \, F=1 \!\rightarrow F^\prime =2$ transition through the atoms along the guiding direction. The image of the probe cross-section and the absorbed atoms are taken by a fast CCD digital camera. Figure~\ref{moving}(b) shows a series of images for different values of a current in two sets of coils: $I_1=50$~A and $I_2 = 0-10$~A (from top to bottom);  the respective guide translation is $\sim 2$~mm (as limited by the cross-section of the probe). The maximum translation is limited by the width of the tape conductors; in our case it can reach over 1 cm according to the current switching algorithm as shown in the simulation in Fig.~\ref{moving}(a).  
The other important application for such a guide would be heat-free coherent transport of guided atoms to a surface for atom-surface interaction studies~\cite{surface}. To transport atoms over several cm distance a multiple-conductor tape can be used.

\begin{figure}[hb!]
\begin{center}
\includegraphics[width=3.2 in]{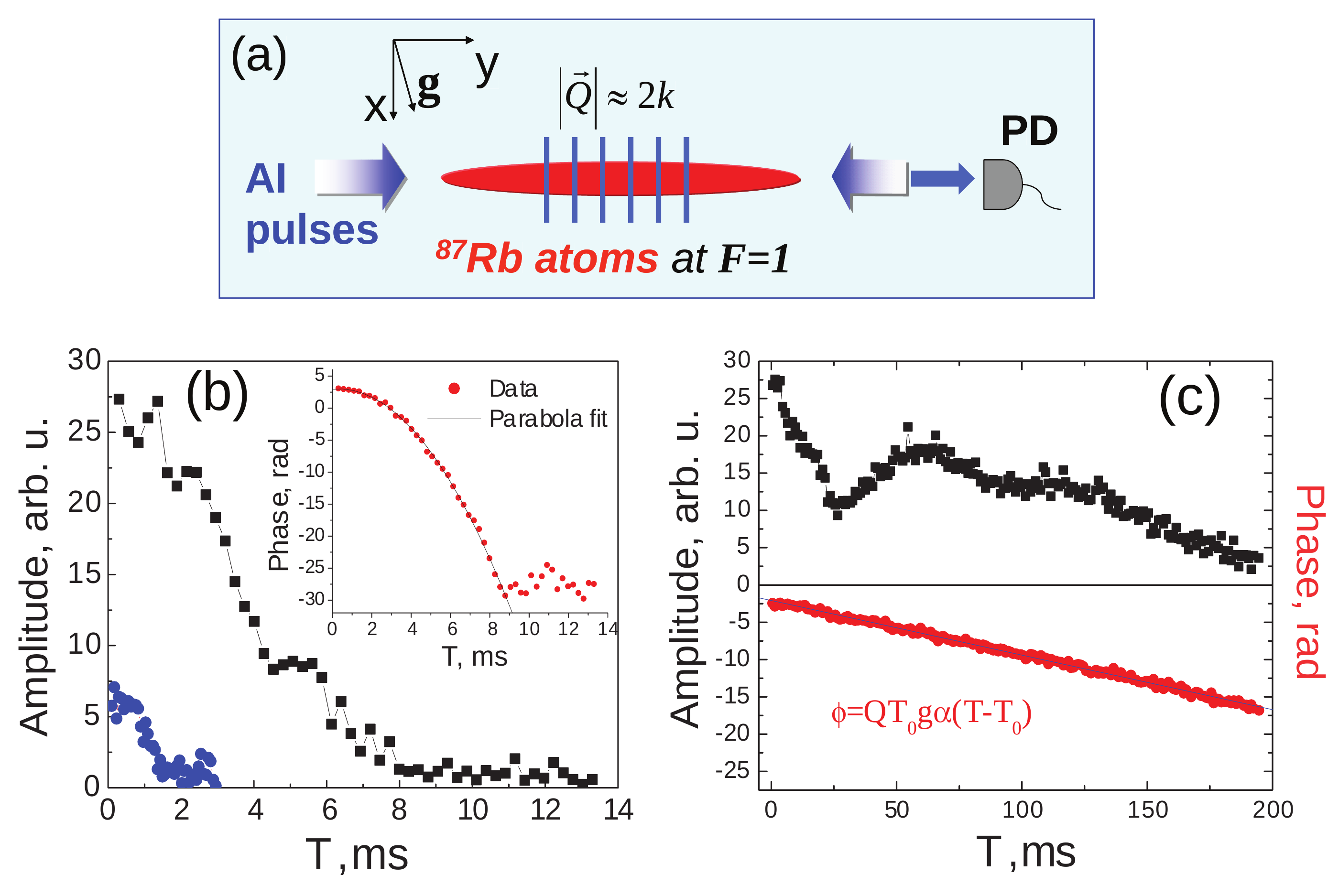}
\end{center}
\caption{(Color online) (a) Atom interferometer (AI) experimental scheme (here, PD is a photodiode, $Q$ is a grating vector); (b) experimental results with three pulse scheme in the symmetric magnetic guide: contrast of the interferometer vs. interrogation time; the inset shows the phase component, which is proportional to the gravitational acceleration component along the guiding direction; (c) experimental results with four-pulse scheme; the formula gives the linear fit, where $T_0$, $\alpha$, {\it g} are the time separation between first and second pulses, the guide tilt to horizon, and gravitational acceleration, respectively.}
\label{data}
\end{figure}

To demonstrate an application of the guided atoms in the macro-guide we carried out an atom interferometer experiment in a time domain similar to the free-space interferometer~\cite{Cahn} and an interferometer with ferro-foils-guided atoms~\cite{Wu}. The interferometer scheme and preliminary results are presented in Fig.~\ref{data}. 
%%%%%%%%%%%%%%%%%%%%%%%%%%%%%%%%%%%%% Experiment scheme
%%%%%%%%%%%%%%%%%%%%%%%%%%%%%%%%%%%%%%%%%%%%%%%%%%%%%%%%%%%% 
The interferometer signal is detected by a heterodyne technique as a function of the interrogation time $T$. Figure~\ref{data}(b) shows a single run data of a contrast decay (an amplitude) and a phase shift for the case when three off-resonant optical standing wave pulses were used as diffraction gratings and Fig.~\ref{data}(c) shows the results of a four-pulse scheme~\cite{Ton08}. In the first case the coherence is preserved for about 9 ms and in the second case for more than 200 ms with the actual wave-packet separation of $d=0.5 \, \mu$m. 
A small component of gravity along the guiding direction contributes to the quadratic phase shift [the inset of Fig.~\ref{data}(b)] and the linear phase for the four-pulse scheme [Fig.~\ref{data}(c)]. As a reference, we also present the typical results of three pulse interferometry experiments with a regular configuration of the macroscopic electromagnet guide as shown by blue circles in Fig.~\ref{data}(b); the symmetric guide provides larger and longer decay of the interferometer contrast. 
The coherence time is highly affected by the misalignment of the optical beams with respect to the guiding axis. The data presented here was obtained using only mechanical stages to align the guiding axis with respect to the optical beams. We hope to increase the coherence time by improving the design of the mounting structure and by implementing high resolution optical beam steering for the alignment. 

In summary, we developed a straight and smooth macroscopic magnetic guiding structure for cold atoms that is a promising tool for atom optics and atom interferometry experiments. The guide can also be used for coherent transport for surface probing and nonlinear optics experiments with cold atoms. 
%The future work would include engineering efforts on improvements of the structure design for better alignment precision. 
%We also plan to develop a ferromagnetic version of symmetric guide. 

%%%%%%%%%%%% END %%%%%%%%%%%%%%%%%%%%%%%%%%%%%%%%%%%%%%%%%%%%%
%\begin{acknowledgement}
We thank T. Kodger for technical assistance. 
This work was supported by the DARPA from DOD, ONR and U.S. Department of the
Army, Agreement Number W911NF-04-1-0032, and by the Draper Laboratory. 
%\end{acknowledgement}
%%%%%%%%%%%%%%%%%%%%%%%%%%%%%%%%%%%%%%%%%%%%%%%%%%%%%%%%%%%%

\end{document}